\documentclass[10pt,twocolumn]{article} 
\usepackage{simpleConference}
\usepackage{times}
\usepackage{graphicx}
\usepackage{amssymb}
\usepackage{url,hyperref}
\usepackage{fancyhdr}			% Para encabezados y pie de pagina
\usepackage{enumitem}
\usepackage{color}
\usepackage{hyperref}
\usepackage{array}

\begin{document}
    \title{\MakeUppercase{COMPARATIVE ANALYSIS USING CLASSIFICATION METHODS VERSUS EARLY STAGE DIABETES}}
    
    \author{
    \begin{tabular}{c}
    Alca-Vilca Gabriel Anthony \\
    Faculty of Statistic and Computer Engineering, \\
    Universidad Nacional del Altiplano de Puno, P.O. Box 291 \\
    Puno - Peru. \\
    Email: galcav@est.unap.edu.pe \\
    \end{tabular}
    \and
    \begin{tabular}{c}
    Carpio-Vargas Eloy \\
    Faculty of Statistic and Computer Engineering, \\
    Universidad Nacional del Altiplano de Puno, P.O. Box 291 \\
    Puno - Peru. \\
    Email: ecarpiov@unap.edu.pe \\
    \end{tabular}
    }
    
    \maketitle
    \thispagestyle{empty}
    \textbf{Abstract—In this research work, a comparative analysis was carried out using classification methods such as: Discriminant Analysis and Logistic Regression to subsequently predict whether a person may have the presence of early stage diabetes. For this purpose, use was made of a database of the UC IRVINE platform of the year 2020 where specific variables that influence diabetes were used for a better result. Likewise in terms of methodology, the corresponding analysis was performed for each of the 3 classification methods and then take them to a comparative table and analyze the results obtained. Finally we can add that the majority of the studies carried out applying the classification methods to the diseases can be clearly seen that there is a certain attachment and more use of the logistic regression classification method, on the other hand, in the results we could see significant differences in terms of the 2 classification methods that were applied, which was valuable information for later drawing final conclusions.}\\
    
    
    \renewcommand\thesection{\Roman{section}} % Numeración en números romanos para las secciones
    \section{INTRODUCTION}
    In the field of health, the search for alternatives to medicine to help medical diagnosis has become a priority. To provide these alternatives, it is necessary to develop fast and accurate information processing systems, i.e. data mining.\textbf{\color{blue}{\cite{silvaidentificaccao}}} On the other hand, in recent years it has been seen how progress in data mining and machine learning has been of great help in the diagnosis of diseases through classification methods. \\
    
    Diabetes today is a very serious disease that affects thousands of people around the world, so early detection of diabetes is very important because we can avoid serious complications and also improve the quality of life of people.\textbf{\color{blue}{\cite{hernandez2021analisis}}}\\
    
    According to the IDF (International Diabetes Federation) reported that by 2021 there will be a total of 537 million adults (20-79) with diabetes and by 2030 it is expected that there will be a total of 643 million people living with diabetes.\textbf{\color{blue}{\cite{inter}}} \\
    
    Excess weight is a well-established risk factor for diabetes. This disease is 5 to 6 times more frequent in obese people than in those with normal weight. The American Diabetes Association has included overweight as one of the risk factors for the detection of Diabetes. The influence of obesity on the risk of developing diabetes is determined not only by its degree, but also by the site where the fat accumulates, for which several mechanisms have been proposed but remain uncertain. \textbf{\color{blue}{\cite{coniglio2020indices}}}\\
    
    Although obesity is strongly related to the existence of IR, not all overweight or obese individuals are insulin-resistant and will be at risk for developing Diabetes; however, it is important to identify who are insulin-resistant. To this end, it was noted that IR estimated by BMI should be complemented with visceral fat distribution determined through waist circumference (WC). Early detection of IR is important for the prevention of clinical manifestations (prediabetes state) that precede Diabetes by several years; detection and control prevent/delay its transition to Diabetes and improve quality of life. \textbf{\color{blue}{\cite{alkhedaide2021association}}}\\
    
    Although the most common method of diagnosing the disease is through inexpensive blood tests, this is not sufficient for a complete assessment. Therefore, this research focuses on finding a non-invasive diagnosis for diabetes through symptoms that may accompany the disease. For this purpose, machine learning algorithms have been implemented on a database of patients with diabetes. \textbf{\color{blue}{\cite{garciahacia}}}\\

    In this study, we focus on two widely used classification methods: discriminant analysis and logistic regression and compare their prediction of the presence of early stage diabetes. Laboratory tests, although strong indicators of pathology, can be misinterpreted. For this reason, a more accurate way to identify patients with the disease has been sought in recent years. \\

    In the following, we will see the development and comparative analysis that was done by applying the classification methods to our database and their subsequent prediction over a period of time.

    \section{DATASETS}
    For the comparative analysis that we will see below by applying the classification methods we focus on the topic of diabetes which is a disease that occurs daily alarmingly in people.\\

    For this analysis we will use the database of IC IRVINE (Machine Learning Repository)\textbf{\textcolor{blue}{\url{https://archive.ics.uci.edu/dataset/529/early+stage+diabetes+risk+prediction+dataset}}}\textbf{\color{blue}{\cite{datos}}} on diabetes which has a total of 520 records and 17 variables of which doing a deep analysis we will focus on 7 variables which are:

    \begin{enumerate}   [label=\arabic*.]
        \item {\textbf{Weight loss: } (Perdida de peso) If the person has lost weight.}
        \item {\textbf{Weakness: } (Debilidad) If the person feels weak, frail, or fragile.}
        \item {\textbf{Irritability: } (Irritabilidad) If the person is easily upset, easily irritated.}
        \item {\textbf{Delayed healing: } (Cicatrizacion retardada) If your wounds do not heal in a timely manner.}
        \item {\textbf{Alopecia: } (Alopecia) Abnormal hair loss.}
        \item {\textbf{Obesity: } (Obesidad) If the person is overweight according to his size.}
        \item {\textbf{Class: } (Clase) See if the person has Diabetes or not.}
    \end{enumerate}
    \begin{figure}[htb]
      \centering
      \includegraphics[width=3.3in]{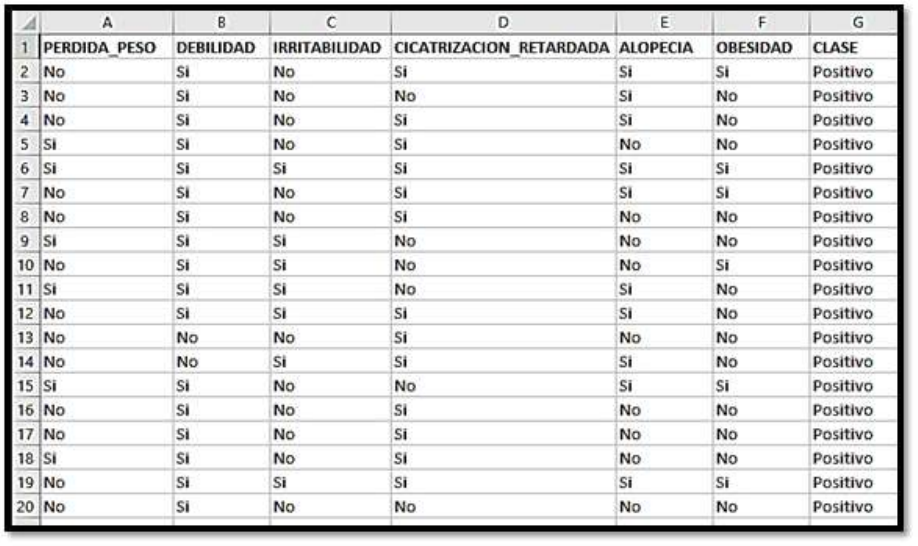}
      \caption{\small \textbf{Early stage diabetes The first 20 records of the entire database are displayed.}}
      \label{fig:figura1}
    \end{figure}

    \section{\MakeUppercase{ materials and methods}}
    \begin{enumerate}[label=\textit{\Alph*)}]
        \item Type of study \\ 
        \\ For this study, an exhaustive review was made of the application of classification methods to medical diseases in recent years. The studies are published in English and Spanish and have been performed worldwide.
      
        \item Techniques and instruments \\
        \\ For this analysis, the data classification technique was used to better analyze the results obtained using the classification method discriminant analysis and logistic regression to our database on early stage diabetes.
       
        \item Procedure and analysis by methods \\
        \\ For this comparative analysis of classification methods, a database of patients with and without early stage diabetes was obtained, including other significant variables that help in the development of the analysis. Also this database was worked separately in the 2 different classification methods divided into 70\% training and 30\% testing using machine learning selected in RStudio that we will see below in detail each method.\\

        For the development of this case we will focus on prediction model and performance metrics where clearly a threshold value is used to assign probability values, when it is greater than 0.5 the result is positive, otherwise it will be negative. The function that relates the dependent variable to the independent variables is called sigmoid function, which is an S-shaped curve that can take any value between 0 and 1, but never values outside these limits. \textbf{\color{blue}{\cite{aparicio2022prediccion}}}\\
        
        Next, we will see 2 validation metrics with which we will discuss once the analysis of this article has been carried out\\ 
        
        ROC curve: statistical tool used to classify individuals in a population into two groups (one representing an event of interest and the other not). Given by the area under the curve (AUC), a standard accuracy metric for binary classification models which measures the ability to predict positive versus negative events, it returns a decimal value between 0 and 1; values close to 1 indicate a very accurate machine learning model. \textbf{\color{blue}{\cite{Javier2018}}} \\
        \begin{figure}[htb]
          \centering
          \includegraphics[width=3.1in]{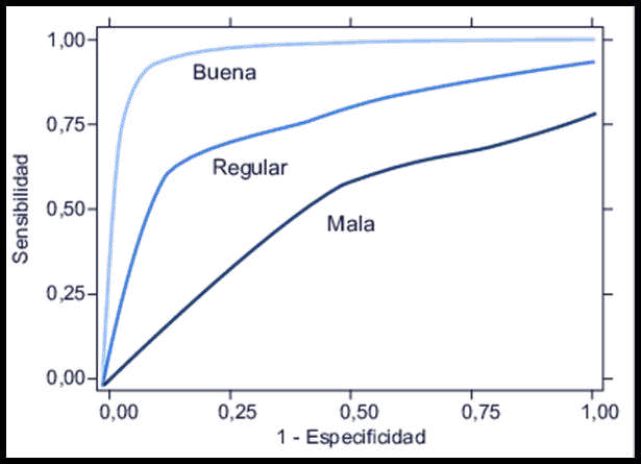}
          \caption{\small \textbf{ROC curve}}
          \label{fig:figura1}
        \end{figure}

        Confusion matrix. It is a tool that allows to obtain the performance of an algorithm, it is applied in binary classification problems (2 classes). It is composed of true positives (VP), false negatives (FN), i.e. cases that were actually positive but the model classified them as negative, false positives (FP) and true negatives (VN) which are cases that were actually negative but the model classified them as positive. \textbf{\color{blue}{\cite{unknown}}} \\
        \begin{figure}[htb]
          \centering
          \includegraphics[width=3in]{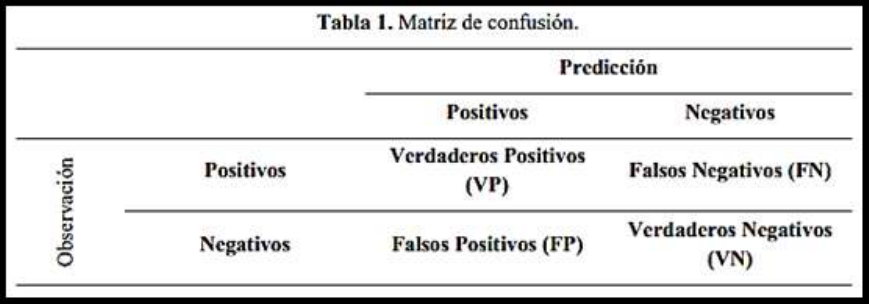}
          \caption{\small \textbf{Confusion matrix}}
          \label{fig:figura1}
        \end{figure}\\ \\

        \begin{enumerate}   [label=\arabic*.]
        \item {\textbf{Discriminant Analysis Method}}\\
            \\This supervised classification method seeks to find a linear combination of predictor variables that maximizes the separation between classes. This is why it works well when the classes are linearly separable and the other variables of the model are satisfied. \\
            
            In the following diagram you can visualize how the classification by variable is given together with the class, this to have an idea of how the data is being handled.\\ 
            \begin{figure}[htb]
              \centering
              \includegraphics[width=2.8in]{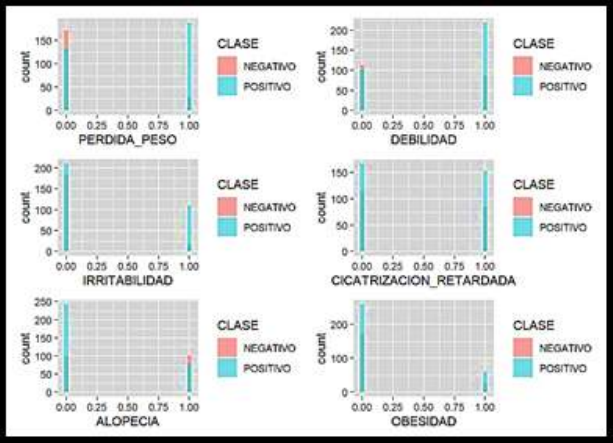}
              \caption{\small \textbf{Graph of the predictor class with each of the variables.}}
              \label{fig:figura1}  
            \end{figure} \\

            Next, it shows us the predictor model, with which we can see, work and analyze the possible predictions for the future.\\ \\ \\ \\ \\ \\ \\ \\ 
            
            \begin{figure}[htb]
              \centering
              \includegraphics[width=2.8in]{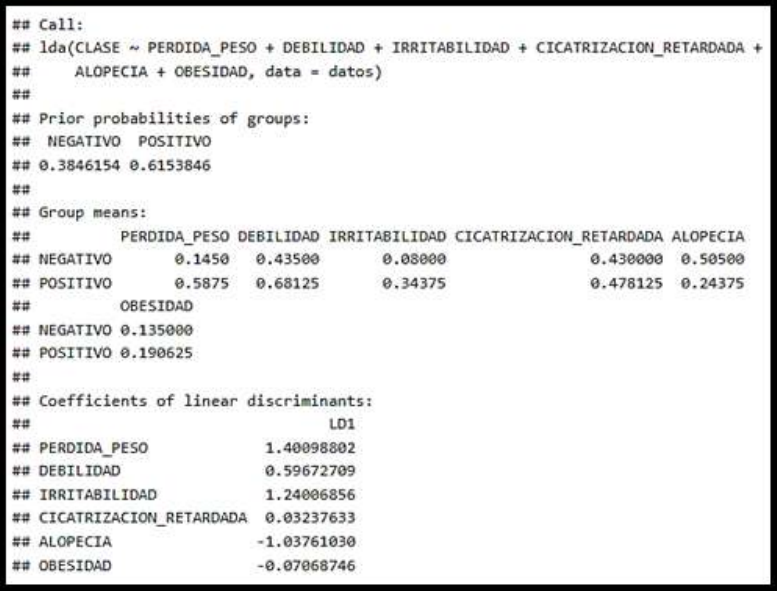}
              \caption{\small \textbf{Predictor model}}
              \label{fig:figura1}  
            \end{figure} 

            One of the first validation metrics that we obtain after having performed an in-depth analysis is this confusion matrix, which indicates how the classification was given once the method was applied to the database entered, giving 0.73\% of no change in the negative classification, 0.26\% of change from negative to positive classification, 0.78\% of no change in the positive classification and 0.21\% of change from positive to negative classification. \\
            \begin{figure}[htb]
              \centering
              \includegraphics[width=2.7in]{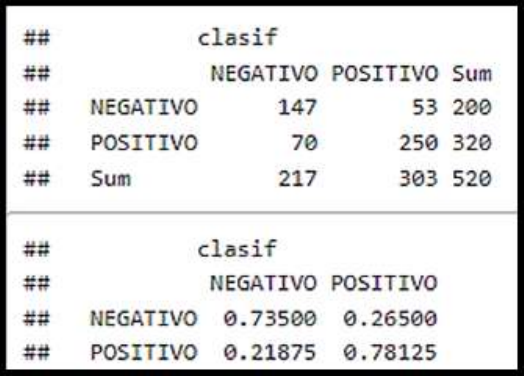}
              \caption{\textbf{Confusion matrix of Discriminant Analysis Method}}
              \label{fig:figura3}   
            \end{figure} 

            The second validation metric that we will see next is the ROC curve, which yields an area of 0.837, indicating that it would validate at 83\% good.\\ \\ \\ \\ \\
            \begin{figure}[htb]
              \centering
              \includegraphics[width=2.7in]{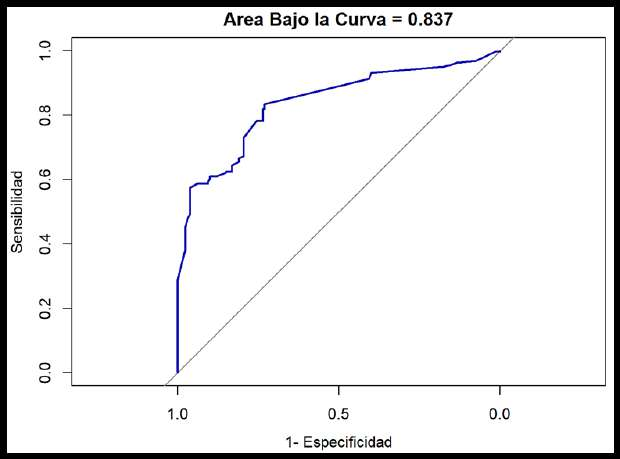}
              \caption{\textbf{ROC curve of Discriminant Analysis Method}}
              \label{fig:figura3}   
            \end{figure}

            Finally, applying the predisposed model we see a prediction of 20 following records based on the given events, we could observe the following predictions, thus giving that of the 20 predictions 14 were positive and 6 were negative\\ 
            \begin{figure}[htb]
              \centering
              \includegraphics[width=2.7in]{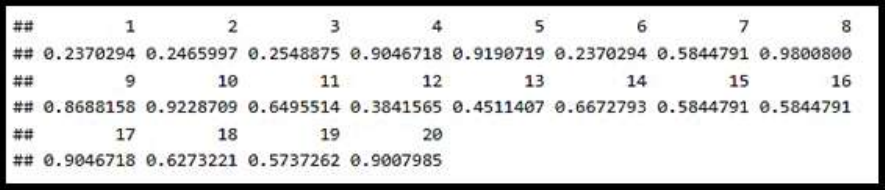}
              \caption{\small \textbf{Predictions of Discriminant Analysis Method}}
              \label{fig:figure4}   
            \end{figure} 

            \item {\textbf{Logistic regression method }} \\ \\
            According \textbf{\color{blue}{\cite{malpartida2022prediccion}}} in his developed paper he indicates that they used the logistic regression method where it is worth noting that this is the most basic and widely used Machine Learning model for binary classification, which can be easily extended to multi-label classification problems. The logistic regression technique uses the sigmoid function to build a regression model that predicts the likelihood that an entry belongs to a particular category. It uses the logistic function to model the probability of belonging to a class, is effective in binary classification problems, and can be extended to multi-class classification problems. However, it assumes a logarithmic relationship between the predictor variables and the response variable.\\
            
            Below, we can see on a large scale how the diabetes variable is presented once the discriminant analysis method has been applied, where we see that 62\% according to the data have diabetes while the other 38\% do not have diabetes.\\ 
            \begin{figure}[htb]
              \centering
              \includegraphics[width=2.8in]{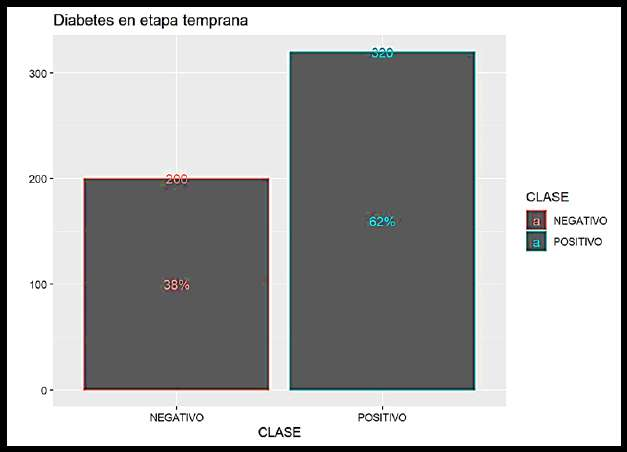}
              \caption{\small \textbf{bar chart of the diabetes variable with the predictor variable}}
              \label{fig:figure4}   
            \end{figure} 
            \begin{figure}[htb]
              \centering
              \includegraphics[width=2.7in]{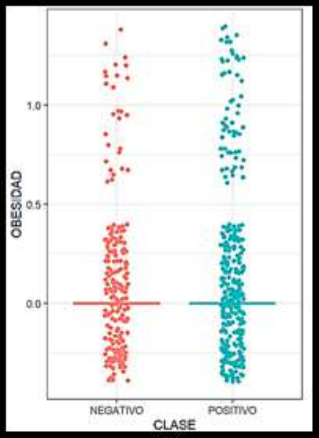}
              \caption{\small \textbf{}}
              \label{fig:figure4}   
            \end{figure} 

            Next, it shows us the predictor model, with which we can see, work and analyze the possible predictions for the future. \\ \\ \\ \\ \\
            \begin{figure}[htb]
              \centering
              \includegraphics[width=2.7in]{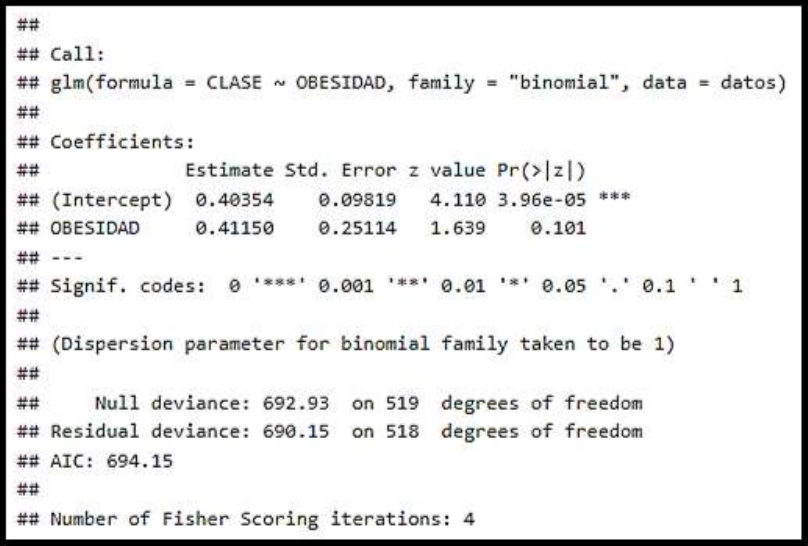}
              \caption{\small \textbf{Predictor model of Logistic regression method}}
              \label{fig:figure4}   
            \end{figure} 
            
            One of the validation metrics presented below is the ROC curve, which shows an area of 0.527, indicating that the validation would be 52\% regular.
            \begin{figure}[htb]
              \centering
              \includegraphics[width=2.7in]{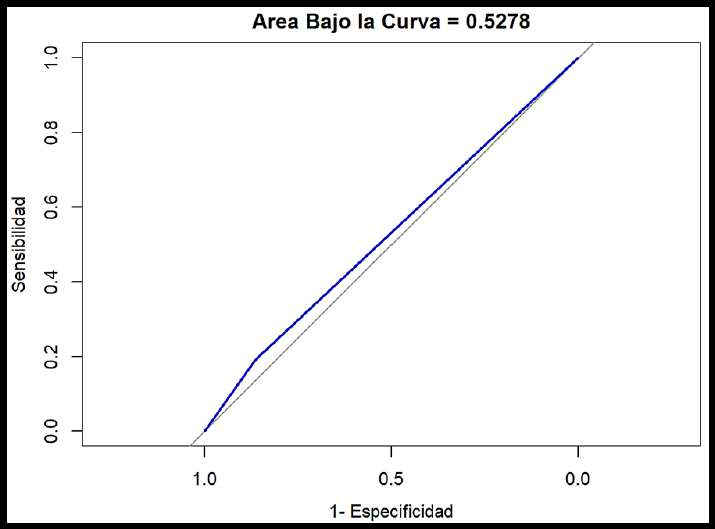}
              \caption{\small \textbf{ROC curve of Logistic regression method}}
              \label{fig:figure4}   
            \end{figure} 

            Finally applying the predisposed model we see a prediction of 20 following records based on the given events we could observe the following predictions, thus giving that of the 20 predictions 20were positive and 0 negative. \\ \\
            \begin{figure}[htb]
              \centering
              \includegraphics[width=2.7in]{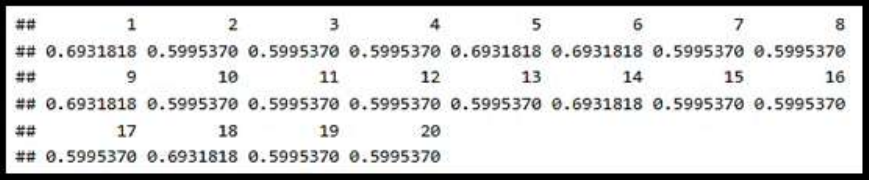}
              \caption{\small \textbf{Predictions of Logistic regression method}}
              \label{fig:figure4}   
            \end{figure} 
            
        \end{enumerate}
    \end{enumerate} 

    \section{ RESULTS} 
    The result after having applied the comparative analysis to the diabetes database revealed that the method that can be used to make predictions is the discriminant analysis method; however, this classification of the different methods differs in terms of the database with which the process is being carried out. \\ \\
    \begin{figure}[htb]
      \centering
      \includegraphics[width=3.4in]{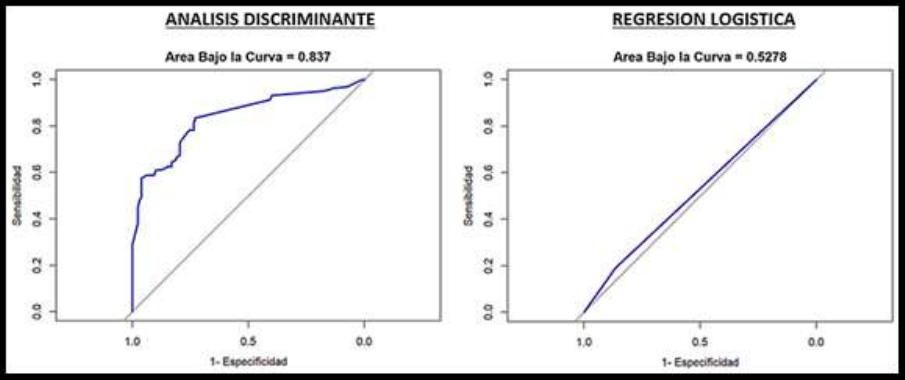}
      \caption{\small \textbf{ROC curve of Discriminant Analysis Method and Logistic regression method}}
      \label{fig:figure4}   
    \end{figure} 

    Clearly in the previous image we can see how the ROC curve, i.e. our validation metric for the two models applied, clearly indicates that the discriminant analysis method has a significant difference compared to the logistic regression method, with 83\% validation in the discriminant analysis and 52\% in the logistic regression. \\ \\

    In \textbf{\color{blue}{\cite{parga2004analisis}}}, a discriminant analysis was performed with the items that had shown significant correlation with the criterion variable, with the aim of establishing a discriminant function from which patients could be classified, thus giving a lambda value of 76\% and its correlation of 65\% for the next predictions.\\ \\

    On the other hand, analyzing the classification we see the following chart, where we can see how the next predictions could be given in a period of 15 sequences, which, working at 95\% confidence, according to the discriminant analysis method and the logistic regression method, show a variation of 31\% of success according to the ROC curve to know if the person could have diabetes or not.\\ \\ \\ \\ \\  \\

    \begin{figure}[htb]
      \centering
      \includegraphics[width=2.7in]{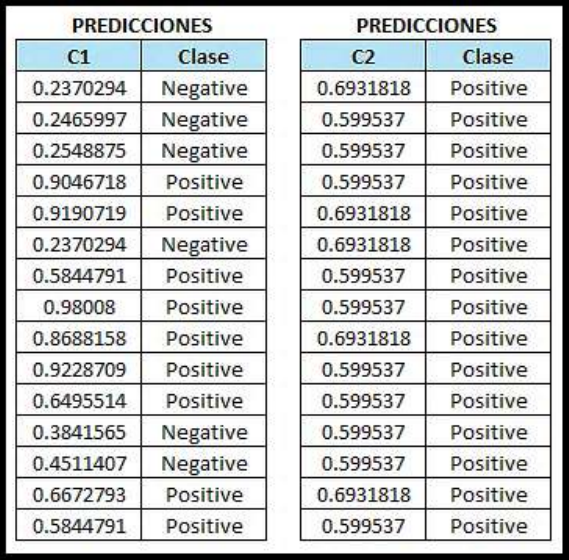}
      \caption{\small \textbf{Predictions of Discriminant Analysis Method and Logistic regression method}}
      \label{fig:figure4}   
    \end{figure} 
    The differences that we can appreciate in more detail are that of the 15 future predictions in the discriminant analysis method, 9 were positive and 6 were negative, and finally in the logistic regression method, all 15 predictions were positive, which indicates that this is the method with which this database should be worked on, since it is the method that was closest in terms of future predictions.\\ \\

    According to \textbf{\color{blue}{\cite{sanchez2021prediccion}}} after performing the respective analysis using the regression method, it yields a Cox and Snell R-squared value (0.552), which indicates that, considering the variables analyzed, 55.2\% can be predicted. This result is very similar to our case study, which would indicate that this method is not as effective for the next predictions. \\ \\

    Likewise, in the following table we can appreciate and make a brief comparison of a few more authors where we can see how their results were applying one of the two classification methods, either the discriminant analysis classification method or the logistic regression classification method. \\ \\ \\ \\ \\ \\ \\ \\ \\

    \begin{table}[t]
        \begin{tabular}{ | m{1.8cm} | m{6cm} | }
        \hline \centering AUTHORS & CONTRIBUTION \\ \hline
        Dogantekin et al. (2010) & They proposed a linear discriminant analysis for the detection of diabetes. A set of variables similar to those used in this work and additional variables were used. With a level of accuracy: 84.61\%. \textbf{\color{blue}{\cite{castrillon2017sistema}}} \\ \hline
        Mario Sequeda (2023) & Taking into account all the results and obtaining similar results in each of the 3 methods, it indicates that the discriminant analysis was more accurate than the other two models. With a level of accuracy: 78\% \textbf{\color{blue}{\cite{diabetes1}}} \\ \hline
        Jaime Rosales (2022) & Several approaches have been performed for classification and early prediction of diabetes using artificial intelligence methods, where by analyzing different medical attributes predict diabetes using five different types of Machine Learning algorithms, showing that LDA and decision tree algorithms outperformed and achieved higher accuracy compared to random forest, logistic regression and Naive Bayes. \textbf{\color{blue}{\cite{malpartida2022prediccion}}} \\ \hline
        \end{tabular}
    \end{table} 

    On the other hand, according to the author Silvio Patricio in his research, he corroborates that as a result it is observed that the models obtained an accuracy ranging between 60\% and 80\% which would indicate that a person would suffer from diabetes. \textbf{\color{blue}{\cite{dos2019classifier}}}\\

    Likewise, according to the results of the author Lopez Raul, two models were obtained that behaved similarly based on the comparison criteria considered for this purpose: percentage of correct classification, sensitivity and specificity. Thus indicating that the model with the automatic interaction detector algorithm using Chi-square was the one with the best predictive results. \textbf{\color{blue}{\cite{lopez2016modelo}}}

    \section{ DISCUSSION AND CONCLUSIONS}
    Comparative analysis of our diabetes database indicated that the logistic regression method did not outperform the discriminant analysis classification method in predicting the presence of early-stage diabetes. Likewise, the high accuracy, sensitivity and specificity achieved by logistic regression do not indicate its potential as an effective tool for the early diagnosis of diabetes.  Therefore, the discriminant analysis method is given as one of the methods to be taken into consideration because of its high demand on the ROC curve presented above. \\

    However \textbf{\color{blue}{\cite{angelucci2021adherencia}}}, indicates that, having worked with the logistic regression method, although the model was significant and allows us to detect variables that predict and correctly classify the group of diabetic patients, the explained variance of the phenomenon did not exceed 50\%, so there is a set of biological, psychological, economic and social factors that can be included.\\

    Therefore, these results suggest the importance of considering different classification methods when developing support systems for the early detection of chronic diseases such as diabetes. On the other hand, further research and clinical studies are needed to further validate and improve the performance of these methods in real clinical settings.\\

    A comment by author Rojas Rosalba indicates that the aging of the population, the insufficiency of screening actions and the increase in diabetes complications will lead to an increase in the burden of disease. Investing in primary and secondary prevention of diabetes is crucial. \textbf{\color{blue}{\cite{rojas2018prevalencia}}} \\

    Finally, according to Dr. Emma Dominguez, in her research, an approach to logistic regression was made, as one of the most widely used multivariate statistical techniques in recent decades, in order to recommend its correct use. Practical issues such as the number of subjects necessary for its application, the situations in which it should be used, the type of variables to which it can be applied, how to include them in the model, the interpretation of the results, etc., were taken into account. An example of the application of this technique in the field of Diabetes was presented. It was concluded that the application of logistic regression is very useful in any field of medical research when we need to determine the effect of a set of variables, potentially considered as influential, on the occurrence of a certain process. \textbf{\color{blue}{\cite{dominguez2001regresion}}} \\

\bibliographystyle{ieeetr}
\bibliography{refs}      
\end{document}